# Incorporating Gas Pipeline Leakage Failure Modes in Risk Evaluation of Electricity-Gas Integrated Energy Systems

Yi Tang, Yuan Zhao*, Wenyuan Li, Kaigui Xie, Juan Yu

***Abstract*—In the existing literatures for the risk evaluation of electricity-gas integrated energy system (EGIES), the impacts of gas leakage in pipelines are ignored. This paper presents a method to incorporate the failure modes of gas pipeline leakage in EGIES risk evaluation. A Markov state transition model of gas pipeline with multi-state and multi-mode transition process, and a bi-level Monte Carlo sampling method for this model are developed. A stochastic topology change based network model of EGIES considering the pipeline leakage failure modes is presented. The risk indices for EGIES based on the load shedding, including those specifically for gas leakage risks, are also proposed. An EGIES with a modified RBTS and a 7-node gas system was used to demonstrate an application of the proposed method and models. The results indicate that pipeline leakage failures have significant impacts on the risk of EGIES. Ignoring pipeline leakage failures in the risk evaluation of EGIES will result in an overly underestimation of system risk and most likely a misleading conclusion in system planning.**

## NOMENCLATURE

*Variables*:

| | |
|---|---|
| $\boldsymbol{\Pi}$ | Vector of gas pressure squared at gas nodes. |
| $\boldsymbol{S}_{\text{st}}$ | Vector of gas discharging from gas storages. |
| $\boldsymbol{S}_{\text{cgc}}$ | Vector of gas consumption in gas compressors. |
| $\boldsymbol{S}_{\text{gc}}$ | Vector of gas flow through gas compressors. |
| $\boldsymbol{S}_{\text{p}}$ | Vector of gas flow through gas pipelines. |
| $\boldsymbol{S}_{\text{gw}}$ | Vector of gas supply from gas wells. |
| $\boldsymbol{S}_{\text{L}}$ | Vector of gas load at gas nodes. |
| $\boldsymbol{S}_{\text{Lc}}$ | Vector of gas load curtailment at gas nodes. |
| $\boldsymbol{S}_{\text{gpg}}$ | Vector of gas consumed by GPGs. |
| $\boldsymbol{S}_{\text{p2g}}$ | Vector of gas generated by P2Gs. |
| $\boldsymbol{P}_{\text{gpg}}$ | Vector of power generated by GPGs. |
| $\boldsymbol{P}_{\text{p2g}}$ | Vector of power consumed by P2Gs. |
| $\boldsymbol{P}_{\text{g}}$ | Vector of power generated by general generators. |
| $\boldsymbol{P}_{\text{w}}$ | Vector of wind power. |
| $\boldsymbol{P}_{\text{wc}}$ | Vector of wind power curtailment. |
| $\boldsymbol{P}_{\text{L}}$ | Vector of electric load at electrical nodes. |
| $\boldsymbol{P}_{\text{Lc}}$ | Vector of electric load curtailment at electrical nodes. |

*Incidence Matrix*:

| | |
|---|---|
| $\boldsymbol{M}^{\text{g}}_{\text{gw}}$ | Gas node-gas well incidence matrix. |
| $\boldsymbol{M}^{\text{g}}_{\text{L}}$ | Gas node-gas load incidence matrix. |
| $\boldsymbol{M}^{\text{g}}_{\text{st}}$ | Gas node-gas storage incidence matrix. |
| $\boldsymbol{M}^{\text{g}}_{\text{p}}$ | Gas node-pipeline incidence matrix. |
| $\boldsymbol{M}^{\text{g}}_{\text{gc}}$ | Gas node-compressor incidence matrix. |
| $\boldsymbol{M}^{\text{g}}_{\text{gpg}}$ | Gas node-GPG incidence matrix. |
| $\boldsymbol{M}^{\text{g}}_{\text{p2g}}$ | Gas node-P2G incidence matrix. |
| $\boldsymbol{M}^{\text{e}}_{\text{TL}}$ | Electrical node-transmission line incidence matrix. |
| $\boldsymbol{M}^{\text{e}}_{\text{gpg}}$ | Electrical node-GPG incidence matrix. |
| $\boldsymbol{M}^{\text{e}}_{\text{p2g}}$ | Electrical node-P2G incidence matrix. |
| $\boldsymbol{M}^{\text{e}}_{\text{g}}$ | Electrical node-general generator incidence matrix. |
| $\boldsymbol{M}^{\text{e}}_{\text{L}}$ | Electrical node-electrical load incidence matrix. |
| $\boldsymbol{M}^{\text{e}}_{\text{w}}$ | Electrical node-wind turbine incidence matrix. |

This work was supported in part by the National Science Fund for Distinguished Young Scholars (51725701) and the National "111" Project of China (Project No. B08036).

Yi Tang, Yuan Zhao, Wenyuan Li, Kaigui Xie and Juan Yu are with the State Key Laboratory of Power Transmission Equipment and System Security and New Technology, Chongqing University, Chongqing, 400044 China. (e-mails: tangyi@cqu.edu.cn; yuanzhao@msn.cn; wenyuan.li@ieee.org; kaiguixie@vip.163.com; yujuancqu@qq.com).

*Operator and Function for Matrix or Vector*:

| | |
|---|---|
| $\boldsymbol{A}\cdot\boldsymbol{B}$ | Element-by-element product of $\boldsymbol{A}$ and $\boldsymbol{B}$. |
| $\boldsymbol{A}\cdot/\boldsymbol{B}$ | Element-by-element division of $\boldsymbol{A}$ and $\boldsymbol{B}$. |
| $\boldsymbol{A}^{\cdot k}$ | Element-wise power of $\boldsymbol{A}$. |
| $(\boldsymbol{A})^{\text{T}}$ | Transpose of matrix $\boldsymbol{A}$. |
| $\lvert\boldsymbol{A}\rvert$ | Compute the absolute value of element in $\boldsymbol{A}$. |
| $sgn(\boldsymbol{A})$ | Compute the sign of element in $\boldsymbol{A}$. |

## I. INTRODUCTION

MORE and more gas-fired power generation stations (GPG) are deployed in power systems because of distinct advantages of GPGs over conventional coal plants [1-3]. At the same time, the emerging technology of power to gas (P2G) links power systems and gas systems together by an energy conversion path in which excessive electricity, such as that from wind power generation, is converted into synthetic natural gas [4]. The wide deployment of GPGs and P2Gs greatly intensify the interaction between the two formerly independent energy infrastructures, developing the concept of electricity-gas integrated energy system (EGIES).

An important topic in this field is the risk assessment of EGIES. Compared with the risk evaluation of power systems alone [5-9], the risk evaluation of EGIES is more challenging due to complex interactions between the two subsystems and the operational characteristics of EGIES. On the one hand, the influence of a certain fault in one subsystem may quickly spread to the other subsystem [10]. For example, once a serious fault happens in the gas supply pipeline of a GPG station, the gas-fired generators will be shut down. On the other hand, the bidirectional energy exchangeability between two subsystems brings beneficial flexibilities for EGIES to deal with accidents. For instance, if some fault happens in the gas network, interrupting the upstream gas supply, gas loads can still be supplied by the connected P2Gs. Therefore, it is necessary to develop the models/methods for the risk evaluation of EGIES.

Many efforts have been made in this regard. The risk assessment of a power system considering only gas supply source reliability was carried out in [11-12]. The reliability of an energy hub without considering constraints of energy networks was assessed in [13-17]. In [18], the availability of energy hub output considering network constraints was evaluated under a given load level. In [19-21], the risk analysis of EGIES was addressed at the energy distribution level. In [22-25], the network constraints were considered for performing a systematic analysis of EGIES reliability.

The failure modes of gas pipelines can be divided into three types: pinhole accident, hole accident and rupture accident [26]. When a severe rupture accident happens to a pipeline due to an external interference or ground movement, the upstream gas cannot pass through the failed pipeline to meet downstream demands. However, when a pinhole or hole accident happens to a pipeline, the upstream gas can still supply the downstream gas load but some unintended gas leaks would be released to the environment until the leak accident is detected and repaired.

Unfortunately, a major deficiency in the existing literatures for the risk evaluation of EGIES is that only rupture failures are considered but the impacts of gas leakages in pipelines, i.e., pinhole and hole leakage failure modes, have been totally ignored. The statistics [26] show that the total occurrence frequency for pinholes and holes is 5-10 times larger than the failure frequency for ruptures. Therefore, the ignorance of the two pipeline gas leakage failure modes will lead to great errors in the evaluation result. Most importantly, the leakages of gas pipelines may cause explosions, fires, and poisoning, resulting in serious casualties, tremendous economic losses, and environmental pollutions. Consequently, a quantitative risk analysis specially for gas leakages in EGIES, which simultaneously considers the amounts, locations and occurrence probabilities of the leakages, is essential in effectively preventing the leakage accidents and reducing the losses caused by the accidents.

It is a challenging task to incorporate the gas pipeline leakage failure modes in the risk evaluation of EGIES. Firstly, the three pipeline failure modes and their transition relations must be represented in a composite model. They have different failure frequencies, durations and consequences. A pinhole accident will evolve into a hole accident and in turn to a rupture accident if the leakage failure is not repaired in time. Secondly, the occurrences, locations, and sizes of the pinhole and hole gas leakages are random and must be randomly sampled. Thirdly, the gas leakage modes must be accurately incorporated into the network formulation for the risk evaluation of EGIES.

To address the challenges mentioned above, this paper makes the following contributions:

1) A four-state Markov state transition model of pipeline is built for representing the multiple failure modes. The statistics provide the frequencies of the three failure modes (but not transition rates) whereas the parameters in the Markov model are the transition rates. A set of non-linear equations are derived and solved to obtain the transition rates by combining the Markov model with the frequency-duration approach.

2) A bi-level Monte Carlo sampling algorithm is proposed to generate a random state transition process from the pipeline Markov multiple state model and locations and sizes of gas leakages. First, a sequential Monte Carlo method is used to sample the duration of current state of each pipeline and form a stochastic Markov Chain for the pipeline multiple state-model. Then, two non-sequential Monte Carlo methods are used in the second level sampling to determine the location and size of the gas leakage if the current state of the pipeline obtained in the first level sampling belongs to one of the two leakage modes.

3) A stochastic topology change based network model for EGIES is presented to minimize the gas load shedding. In this network model, a concept of virtual loads is proposed for simulating the influence of gas leaks. Because the locations of virtual gas load nodes are randomly determined, not only virtual gas loads but also the topology of network will stochastically change.

The paper is organized as follows. The four-state Markov model of pipelines and the calculations for model parameters are detailed in Section II. The bi-level Monte Carlo sampling algorithm for determining the specific gas leakage failure mode of pipelines is developed in Section III. The risk evaluation framework of EGIES with consideration of the impact of gas leakage is presented in Section IV. Case studies are given in Section V to demonstrate the impact of pipeline leakage failure modes on the risk evaluation of EGIES, followed by conclusions in Section VI.

## II. Markov Model of Pipelines with Leakage Failures

The three failure modes of pinhole, hole and rupture accidents for pipelines represent three different states associated with different failure frequencies, durations and consequences. A Markov state space diagram [7] can be used to capture the transitions between them and between the normal state and them, as shown in Fig. 1. State 1 represents the normal state, state 2 the failed or repairing state, state 3 the minor gas leak state, and state 4 the serious gas leak state. The minor gas leak represents a case in which a pipeline has a pinhole accident leading to a small amount of gas leakage, which does not create an immediate safety or security risk and thus is allowed to take a relatively long time to be repaired. The serious gas leak state represents a case in which a pipeline has a hole accident leading to a large amount of gas leakage, which must be treated within a very limited time period. A rupture failure happens so fast to directly reach the repair state. Transition rates $\lambda_{12}$, $\lambda_{13}$, and $\lambda_{14}$ represent that a normal operating pipeline may enter the rupture accident, pinhole accident, and hole accident, respectively. Transition rate $\lambda_{21}$ represents the recovery to the normal state from the failed state. Transition rates $\lambda_{32}$ represents that the pinhole accident is found in a routine patrolling and enters the repair, whereas $\lambda_{34}$ represents that the pinhole accident evolves into a hole accident because of extension of corrosion before a timely interference. Transition rate $\lambda_{42}$ represents that the serious leak is detected and a timely action is taken allowing the pipeline entering the repair.

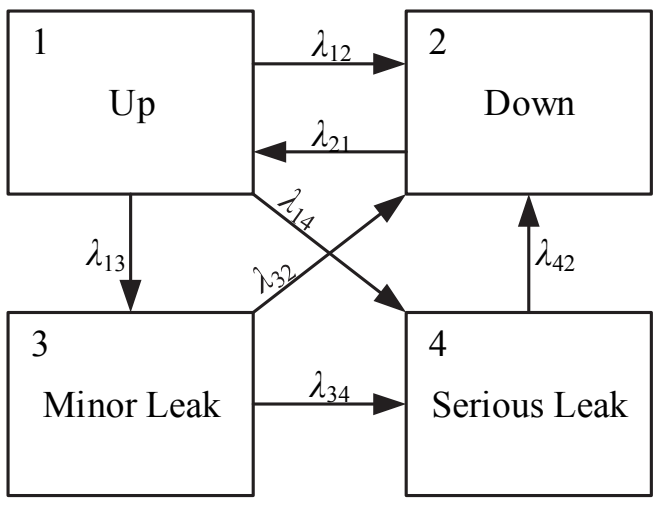

Fig. 1. State space diagram for a pipeline with three failure modes

Once the parameters of the model, i.e., the transition rates, are obtained, a random state transition process can be generated from the state space diagram (see Section III below). Some parameters can be easily obtained from statistics according to their definitions. For example, transition rate $\lambda_{32}$ is equal to the reciprocal of the mean time between two consecutive personnel patrols. Other parameters except for $\lambda_{12}$, $\lambda_{13}$, and $\lambda_{14}$ have similar significances and can be calculated from statistics data. However, statistics do not directly provide the transition rates $\lambda_{12}$, $\lambda_{13}$, and $\lambda_{14}$ but usually only give the frequencies of encountering the three failure accident events. Fortunately, the Markov equation method can be combined with the frequency-duration approach [7] to yield the analytic expression of the transition rates $\lambda_{12}$, $\lambda_{13}$, and $\lambda_{14}$.

Equation (1) can be established from the state space diagram by using Markov equation method, where $P_j$ is the limiting state probability of state $j$.

$$\begin{bmatrix} 1 & 1 & 1 & 1 \\ \lambda_{12} & -\lambda_{21} & \lambda_{32} & \lambda_{42} \\ \lambda_{13} & 0 & -\lambda_{32}-\lambda_{34} & 0 \\ \lambda_{14} & 0 & \lambda_{34} & -\lambda_{42} \end{bmatrix} \begin{bmatrix} P_1 \\ P_2 \\ P_3 \\ P_4 \end{bmatrix} = \begin{bmatrix} 1 \\ 0 \\ 0 \\ 0 \end{bmatrix} \quad (1)$$

The analytic solution of this Markov equation is as follows:

$$P_1 = \frac{1}{\dfrac{\lambda_{13}}{\lambda_{32}+\lambda_{34}} + \dfrac{\lambda_{12}+\lambda_{13}+\lambda_{14}}{\lambda_{21}} + \dfrac{\lambda_{14}+\dfrac{\lambda_{34}\lambda_{13}}{\lambda_{32}+\lambda_{34}}}{\lambda_{42}} + 1} \quad (2)$$

$$P_2 = \frac{\lambda_{12}+\lambda_{13}+\lambda_{14}}{\lambda_{21}} P_1 \quad (3)$$

$$P_3 = \frac{\lambda_{13}}{\lambda_{32}+\lambda_{34}} P_1 \quad (4)$$

$$P_4 = \frac{\lambda_{14}+\dfrac{\lambda_{34}\lambda_{13}}{\lambda_{32}+\lambda_{34}}}{\lambda_{42}} P_1 \quad (5)$$

According to the frequency-duration method [7], the frequency of encountering a state is equal to the product of the probability of this state and the sum of transition rates departing from this state. Therefore, the following equations are obtained:

$$\begin{cases} P_2\lambda_{21} = f_2 \\ P_3(\lambda_{32}+\lambda_{34}) = f_3 \\ P_4\lambda_{42} = f_4 \end{cases} \quad (6)$$

where $f_j$ is the frequency of encountering state $j$.

By substituting Equations (2)-(5) into Equation (6), the following formulas for $\lambda_{12}$, $\lambda_{13}$, and $\lambda_{14}$ can be derived:

$$\begin{cases} \lambda_{12} = \dfrac{\lambda_{21}\lambda_{42}((f_3+f_4-f_2)\lambda_{32}+(f_4-f_2)\lambda_{34})}{(f_2\lambda_{42}+f_4\lambda_{21}-\lambda_{21}\lambda_{42})(\lambda_{32}+\lambda_{34})+f_3\lambda_{21}\lambda_{42}} \\ \lambda_{13} = \dfrac{-f_3\lambda_{21}\lambda_{42}(\lambda_{32}+\lambda_{34})}{(f_2\lambda_{42}+f_4\lambda_{21}-\lambda_{21}\lambda_{42})(\lambda_{32}+\lambda_{34})+f_3\lambda_{21}\lambda_{42}} \\ \lambda_{14} = \dfrac{-\lambda_{21}\lambda_{42}(f_4\lambda_{32}+f_4\lambda_{34}-f_3\lambda_{34})}{(f_2\lambda_{42}+f_4\lambda_{21}-\lambda_{21}\lambda_{42})(\lambda_{32}+\lambda_{34})+f_3\lambda_{21}\lambda_{42}} \end{cases} \quad (7)$$

All the transition rates in the right side of Equation (7) and the frequencies $f_3$ and $f_4$ can be directly obtained from statistics. $f_3$ and $f_4$ are the frequencies of encountering the pinhole accident and hole accident, respectively. $f_2$ is the frequency of encountering the repairing state and the frequency of the rupture accident (denoted by $f_0$) is only a part of it. In case where $f_2$ is not available from a data system, it can be calculated as follows. Based on the frequency balance approach, the frequency of leaving a state is equal to the frequency of entering the state [7, 27]. In this case, the following Equation (8) holds:

$$f_2: \ P_2\lambda_{21} = P_1\lambda_{12} + P_3\lambda_{32} + P_4\lambda_{42} \quad (8)$$

where $P_1\lambda_{12}$ is equal to the frequency of encountering the rupture accident $f_0$, which can be directly collected in statistics. By substituting Equation (6) into (8), $f_2$ can be calculated by:

$$f_2 = f_0 + \frac{\lambda_{32}}{\lambda_{32}+\lambda_{34}} f_3 + f_4 \quad (9)$$

## III. Bi-level Monte Carlo Sampling Algorithm

With the multiple state Markov model of pipelines and all the transition rates in the model, the following bi-level Monte Carlo sampling algorithm is proposed to simulate the multi-state and multi-mode transition process of pipelines and random locations and sizes of gas leakages. For other components in both electricity and gas subsystems, the traditional sequential Monte Carlo simulation method is used.

In the first level, a sequential sampling method is used to sample the duration of current state and transition to next state for the pipeline model. It includes the following steps:

Step 1) Identify the current state in the pipeline model and all the possible transitions departing from the current state.

Step 2) Sample all the possible current state durations according to the transition rates from the current state to possible next states.

Step 3) Select the minimal one among the possible durations as the current state duration and the corresponding state transition to determine the next transited state.

For example, a pipeline is assumed to reside in the up state at the beginning of the simulation. There are three possible states to follow: the down state, the minor leak state, or the serious leak state. If the pipeline transits to the down state, the sampled value of the up-state duration is given by

$$T_{\mathrm{up1}} = -\frac{1}{\lambda_{12}} \ln U_1 \quad (10)$$

If the pipeline transits to the minor leak state, the sampled value of the up-state duration is given by

$$T_{\mathrm{up2}} = -\frac{1}{\lambda_{13}} \ln U_2 \quad (11)$$

If the pipeline transits to the serious leak state, the sampled value of the up-state duration is given by

$$T_{\text{up3}} = -\frac{1}{\lambda_{14}} \ln U_3 \tag{12}$$

where $U_1$, $U_2$, and $U_3$ are three uniformly distributed random numbers between 0 and 1. The up-state duration selected in the Markov chain is:

$$T_{\text{up}} = \min(T_{\text{up1}}, T_{\text{up2}}, T_{\text{up3}}) \tag{13}$$

Equation (13) also indicates the next state of the pipeline. If $T_{\text{up}}$=$T_{\text{up1}}$, the next state is the down state. If $T_{\text{up}}$=$T_{\text{up2}}$, the next state is the minor leak state. If $T_{\text{up}}$=$T_{\text{up3}}$, the next state is the serious leak state.

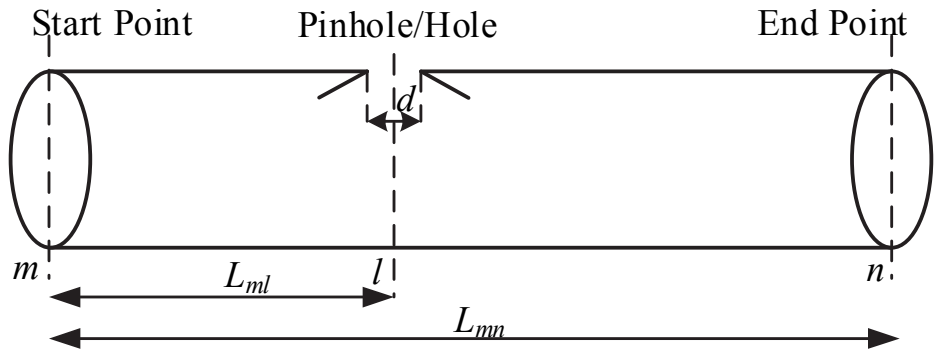

Fig. 2. A pipeline in leakage failure modes

In the second level, two independent random numbers are sampled to decide the location and size of the gas leak if the pipeline is in either leakage state which was determined by the first level sampling. As shown in Fig. 2, for a pipeline with the length of $L_{mn}$, a gas leak happens in the point $l$ that is located at a $L_{ml}$ distance away from the start point $m$, and the diameter of leak hole is $d$ with an assumption of circle shape. The location and size of the leakage are sampled as follows:

Step 1) Sample a uniformly distributed random number $U$ between 0 and 1. The location of the gas leak is decided by the length $L_{ml}$ away from the start point:

$$L_{ml} = L_{mn} U \tag{14}$$

Step 2) Sample a normal distributed random number $d \sim N(\mu,\sigma)$ as the diameter of leak hole, where parameters $\mu$ and $\sigma$ vary according to the type of leakage failure mode. The area $A$ of the hole of gas leak is decided by:

$$A = \frac{\pi}{4} d^2 \tag{15}$$

## IV. Risk Evaluation of EGIES Considering Pipeline Leakage Failure Modes

The incorporation of pipeline leakage failure modes not only requires a new reliability model of pipeline, but also changes the operation manner of pipeline and the network topology in the risk evaluation of EGIES. The first problem has been discussed in Sections II and III. The second problem is discussed in this section. A concept of virtual load is proposed as a representation of gas leak in pipeline. One original pipeline is divided into two separate parts due to introducing a virtual load node, which changes the pipeline network topology and increases the number of nodes in power balance, resulting in a modified network model. Because the number, locations and sizes of virtual load nodes are random in sampling, the network model includes stochastic topology structure and load nodes.

### *A. Energy Flow Equations of Pipeline Considering Leakage Failure Modes*

The gas flow along a pipeline at time t is expressed by [28]:

$$Q_{mn,t} = d_n^m \left( d_n^m \frac{\pi_{m,t}^2 - \pi_{n,t}^2}{R_{mn}} \right)^{0.5}, \text{ where } R_{mn} = \frac{L_{mn} G T_{\text{a}} Z_{\text{a}} f \pi_{\text{b}}^2}{\text{C}^2 T_{\text{b}}^2 D^5 e^2} \tag{16}$$

where $Q_{mn,t}$ is the flow rate (standard cubic meters/day), $\pi_{m,t}$ is the inlet pressure (kPa), $\pi_{n,t}$ is the outlet pressure (kPa), $d_n^m$ is the reference direction coefficient which is equal to 1 if $\pi_{m,t}$ is larger than $\pi_{n,t}$ or -1 if otherwise, and $R_{mn}$ is the hydraulic resistance coefficient where $L_{mn}$ is the length of the pipeline and the definitions of other parameters can be found in [28].

Note that confusions may be caused in the EGIES research due to the different physical units for electrical energy and gas energy. Thus, the unit of gas flow in gas subsystem is converted from standard m$^3$/day into MW by using the energy density of natural gas, i.e. lower heating value ($V_{LH}$) in MJ/m$^3$ [29]:

$$S_{mn,t} = d_n^m \left( d_n^m \frac{\pi_{m,t}^2 - \pi_{n,t}^2}{Z_{mn}} \right)^{0.5}, \text{ where } Z_{mn} = \frac{7464.96 R_{mn}}{V_{LH}^2} \tag{17}$$

where $S_{mn,t}$ is the energy flow rate (MW), $\pi_{m,t}$ is the inlet pressure (MPa), $\pi_{n,t}$ is the outlet pressure (MPa), and $Z_{mn}$ is the resistance coefficient that is proportional to the length of pipe.

Because of gas compressibility and a limited gas flow velocity, the gas flows at both ends of a pipeline are not equal due to a hysteresis effect, and the imbalance of gas flow is stored in or compensated by the stored gas in pipeline. Mathematically, these features can be expressed as [30]:

$$S_{mn,t} = (S_{mn,\text{u},t} + S_{mn,\text{d},t})/2 \tag{18}$$

$$E_{mn,t} = E_{mn,t-\Delta t} + (S_{mn,\text{u},t} - S_{mn,\text{d},t})\Delta t \tag{19}$$

$$E_{mn,t} = K_{mn}\left(\pi_{m,t} + \pi_{n,t}\right)/2 \tag{20}$$

where $S_{mn,\text{u},t}$ and $S_{mn,\text{d},t}$ are the upstream and downstream energy flows of pipeline at time $t$, respectively; $E_{mn,t}$ denotes the gas energy stored in pipe; and $K_{mn}$ is a pipeline pack coefficient which is proportionally to the length $L_{mn}$.

The expressions of upstream and downstream energy flows can be derived from Equations (17)-(20), as shown below:

$$S_{mn,\text{u},t} = d_n^m \left( d_n^m \frac{\pi_{m,t}^2 - \pi_{n,t}^2}{Z_{mn}} \right)^{0.5} + K_{mn} \frac{\left(\pi_{m,t} + \pi_{n,t}\right) - \left(\pi_{m,t-1} + \pi_{n,t-1}\right)}{4} \tag{21}$$

$$S_{mn,\text{d},t} = d_n^m \left( d_n^m \frac{\pi_{m,t}^2 - \pi_{n,t}^2}{Z_{mn}} \right)^{0.5} - K_{mn} \frac{\left(\pi_{m,t} + \pi_{n,t}\right) - \left(\pi_{m,t-1} + \pi_{n,t-1}\right)}{4} \tag{22}$$

When a pipeline transits to a leakage failure state, the energy flow equations (21) and (22) for this pipeline must change due to the appearance of a virtual load node. As illustrated in Fig. 3, the energy flow equations for the segment of the pipeline from start point $m$ to leak point $l$ are presented in (23)-(26), and the energy flow equations for the segment of the pipeline from leak point $l$ to end point $n$ are presented in (27)-(30):

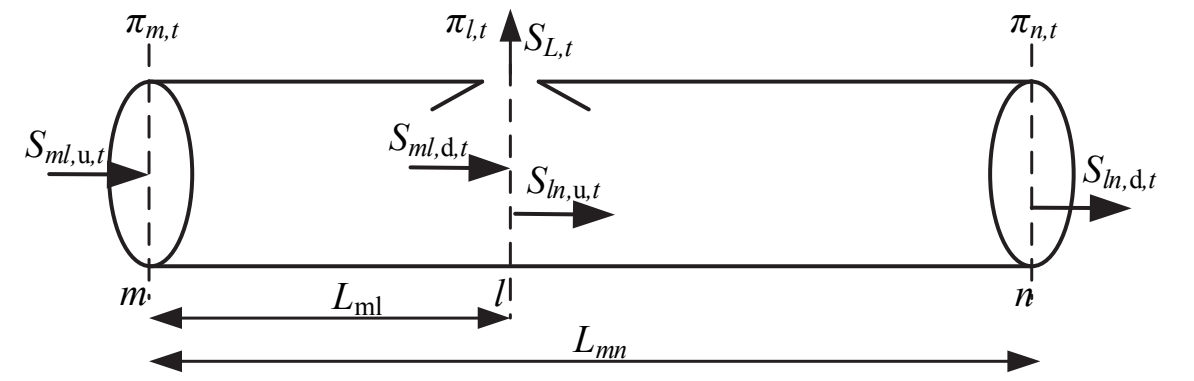

Fig. 3. Schematic illustration of energy flow of pipeline with a gas leak

$$S_{ml,\mathrm{u},t}=d_l^m\left(d_l^m\,\frac{\pi_{m,t}^2-\pi_{l,t}^2}{Z_{ml}}\right)^{0.5}+K_{ml}\,\frac{\left(\pi_{m,t}+\pi_{l,t}\right)-\left(\pi_{m,t-1}+\pi_{l,t-1}\right)}{4} \quad (23)$$

$$S_{ml,\mathrm{d},t}=d_l^m\left(d_l^m\,\frac{\pi_{m,t}^2-\pi_{l,t}^2}{Z_{ml}}\right)^{0.5}-K_{ml}\,\frac{\left(\pi_{m,t}+\pi_{l,t}\right)-\left(\pi_{m,t-1}+\pi_{l,t-1}\right)}{4} \quad (24)$$

$$Z_{ml}=\frac{L_{ml}}{L_{mn}}Z_{mn} \quad (25)$$

$$K_{ml}=\frac{L_{ml}}{L_{mn}}K_{mn} \quad (26)$$

$$S_{ln,\mathrm{u},t}=d_n^l\left(d_n^l\,\frac{\pi_{l,t}^2-\pi_{n,t}^2}{Z_{ln}}\right)^{0.5}+K_{ln}\,\frac{\left(\pi_{l,t}+\pi_{n,t}\right)-\left(\pi_{l,t-1}+\pi_{n,t-1}\right)}{4} \quad (27)$$

$$S_{ln,\mathrm{d},t}=d_n^l\left(d_n^l\,\frac{\pi_{l,t}^2-\pi_{n,t}^2}{Z_{ln}}\right)^{0.5}-K_{ln}\,\frac{\left(\pi_{l,t}+\pi_{n,t}\right)-\left(\pi_{l,t-1}+\pi_{n,t-1}\right)}{4} \quad (28)$$

$$Z_{ln}=\left(1-\frac{L_{ml}}{L_{mn}}\right)Z_{mn} \quad (29)$$

$$K_{ln}=\left(1-\frac{L_{ml}}{L_{mn}}\right)K_{mn} \quad (30)$$

where $\pi_{l,t}$ is the leak point pressure (MPa), and $L_{ml}$ is the length of the segment of pipeline from start point $m$ to leak point $l$.

Additionally, the number of nodes for power balance in the EGIES model increases with the appearance of virtual load nodes. In this case, the increased power balance equation is presented in (31), and the reference direction of power flow is illustrated in Fig. 4.

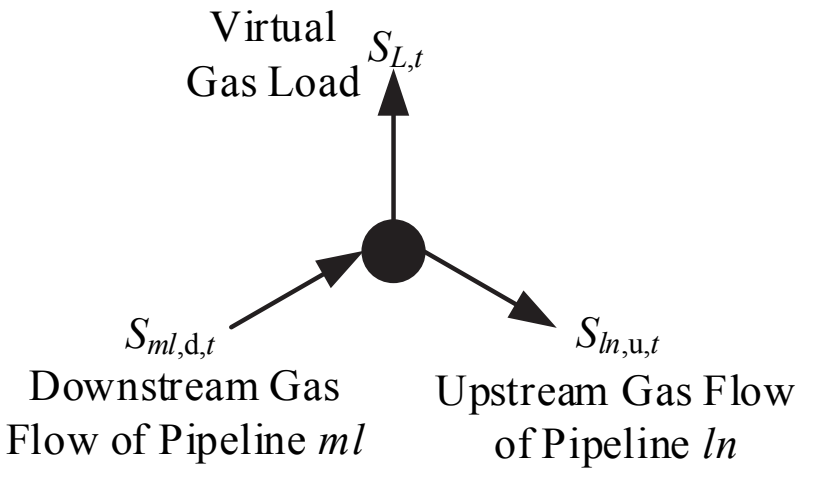

Fig. 4. Reference direction of energy flow for a virtual load node

$$S_{ml,\mathrm{d},t}-S_{ln,\mathrm{u},t}-S_{L,t}=0 \quad (31)$$

where $S_{L,t}$ is the leak power (MW) as shown below [31]:

$$S_{L,t}=A\cdot\pi_{l,t}\cdot\left(\frac{\mathrm{M}\cdot k}{\mathrm{R}\cdot T_{\mathrm{a}}}\left(\frac{2}{k+1}\right)^{\frac{k+1}{k-1}}\right)^{0.5}\frac{V_{LH}}{\rho} \quad (32)$$

where $A$ is the area of leak hole (mm$^2$), $\pi_{l,t}$ is the leak pressure (MPa), M is the molar mass (kg/kmol), R is the gas constant (J/(kmol·K)), $T_{\mathrm{a}}$ is the operation temperature (K), $k$ is the heat capacity ratio, $V_{LH}$ is the lower heating value (MJ/m$^3$), and $\rho$ is the gas density (kg/m$^3$). It is worth noting again that because the number, locations and sizes of virtual load nodes are random, the above equations contain random variables and stochastic topology.

*B. Network Model of EGIES*

*1) Equality Constraints of Gas Subsystem:*

$$\boldsymbol{S}_{\mathrm{p},t}-sgn\left((\boldsymbol{M}_{\mathrm{p}}^{\mathrm{g}})^{\mathrm{T}}\boldsymbol{\Pi}_t\right)\cdot\left(\left|(\boldsymbol{M}_{\mathrm{p}}^{\mathrm{g}})^{\mathrm{T}}\boldsymbol{\Pi}_t\right|./\,\boldsymbol{Z}_{\mathrm{p}}\right)^{\cdot 0.5}=0 \quad (33)$$

$$\boldsymbol{S}_{\mathrm{p,u},t}-\boldsymbol{S}_{\mathrm{p},t}+\boldsymbol{K}_{\mathrm{p}}\cdot\left(\left|(\boldsymbol{M}_{\mathrm{p}}^{\mathrm{g}})^{\mathrm{T}}\right|\boldsymbol{\Pi}_t^{\cdot 0.5}-\left|(\boldsymbol{M}_{\mathrm{p}}^{\mathrm{g}})^{\mathrm{T}}\right|\boldsymbol{\Pi}_{t-1}^{\cdot 0.5}\right)/4=0 \quad (34)$$

$$\boldsymbol{S}_{\mathrm{p,d},t}-\boldsymbol{S}_{\mathrm{p},t}-\boldsymbol{K}_{\mathrm{p}}\cdot\left(\left|(\boldsymbol{M}_{\mathrm{p}}^{\mathrm{g}})^{\mathrm{T}}\right|\boldsymbol{\Pi}_t^{\cdot 0.5}-\left|(\boldsymbol{M}_{\mathrm{p}}^{\mathrm{g}})^{\mathrm{T}}\right|\boldsymbol{\Pi}_{t-1}^{\cdot 0.5}\right)/4=0 \quad (35)$$

$$\boldsymbol{S}_{\mathrm{cgc},t}-\boldsymbol{K}_{\mathrm{gc}}\cdot\boldsymbol{S}_{\mathrm{gc},t}\cdot(\boldsymbol{R}_t^{\cdot\boldsymbol{\alpha}}-1)=0 \quad (36)$$

$$\boldsymbol{R}_t^{\cdot 2}\cdot\left((\mathbf{M}_{\mathrm{gc,u}}^{g})^{\mathrm{T}}\boldsymbol{\Pi}_t\right)-\left((\mathbf{M}_{\mathrm{gc,d}}^{g})^{\mathrm{T}}\boldsymbol{\Pi}_t\right)=0 \quad (37)$$

$$\boldsymbol{E}_{\mathrm{st},t}^{\mathrm{w}}-\boldsymbol{E}_{\mathrm{st},t-\Delta t}^{\mathrm{w}}-\boldsymbol{S}_{\mathrm{st,dis},t}\cdot\Delta t=0 \quad (38)$$

$$\boldsymbol{S}_{\mathrm{st,dis},t}^{\max}-\boldsymbol{K}_{\mathrm{st},1}\cdot\boldsymbol{E}_{\mathrm{st},t}^{\mathrm{w}\,\cdot 0.5}=0 \quad (39)$$

$$\boldsymbol{S}_{\mathrm{st,cha},t}^{\max}-\boldsymbol{K}_{\mathrm{st},2}\cdot\left(1./\left(\boldsymbol{E}_{\mathrm{st},t}^{\mathrm{w}}+\boldsymbol{E}_{\mathrm{st},t}^{\mathrm{c}}\right)+\boldsymbol{K}_{\mathrm{st},3}\right)^{\cdot 0.5}=0 \quad (40)$$

$$\begin{aligned}&\boldsymbol{M}_{\mathrm{gw}}^{\mathrm{g}}\boldsymbol{S}_{\mathrm{gw},t}+\boldsymbol{M}_{\mathrm{st}}^{\mathrm{g}}\boldsymbol{S}_{\mathrm{st},t}-\boldsymbol{M}_{\mathrm{L}}^{\mathrm{g}}(\boldsymbol{S}_{\mathrm{L},t}-\boldsymbol{S}_{\mathrm{Lc},t})-\boldsymbol{M}_{\mathrm{p,u}}^{g}\boldsymbol{S}_{\mathrm{p,u},t}+\boldsymbol{M}_{\mathrm{p,d}}^{\mathrm{g}}\boldsymbol{S}_{\mathrm{p,d},t}\\&-(\boldsymbol{M}_{\mathrm{gc,u}}^{g}-\boldsymbol{M}_{\mathrm{gc,d}}^{g})\boldsymbol{S}_{\mathrm{gc},t}-\boldsymbol{M}_{\mathrm{gc,u}}^{g}\boldsymbol{S}_{\mathrm{cgc},t}-\boldsymbol{M}_{\mathrm{gpg}}^{g}\boldsymbol{S}_{\mathrm{gpg},t}+\boldsymbol{M}_{\mathrm{p2g}}^{\mathrm{g}}\boldsymbol{S}_{\mathrm{p2g},t}=0\end{aligned} \quad (41)$$

where constraints (33)-(35) give the upstream and downstream gas flows of pipelines in the reference direction, $\boldsymbol{S}_{\mathrm{p},t}$ is a vector of energy flow at time $t$, i.e., $\boldsymbol{S}_{\mathrm{p},t}=[S_{12,t}, \ldots, S_{mn,t}, \ldots]^{\mathrm{T}}$, and $\boldsymbol{M}_{\mathrm{p}}^{\mathrm{g}}$ is the gas node-pipeline incidence matrix of which the number of rows is equal to the number of gas nodes, the number of columns is equal to the number of pipelines, and the corresponding element is 1 if the gas node is the start point of a pipeline in reference direction, -1 if the gas node is the end point of that pipeline, or 0 if otherwise. Other incidence matrixes have the similar definitions. Note that if any pipeline is sampled to transit to a leakage failure state, the corresponding gas flow constrains (33)-(35) for that pipeline are replaced by constrains (23)-(32). Constrains (36)-(37) describe the operation characteristics of gas compressors [29], where $\boldsymbol{M}_{\mathrm{gc,u}}^{\mathrm{g}}=1/2(|\boldsymbol{M}_{\mathrm{gc}}^{\mathrm{g}}|+\boldsymbol{M}_{\mathrm{gc}}^{\mathrm{g}})$, $\boldsymbol{M}_{\mathrm{gc,d}}^{\mathrm{g}}=1/2(|\boldsymbol{M}_{\mathrm{gc}}^{\mathrm{g}}|-\boldsymbol{M}_{\mathrm{gc}}^{\mathrm{g}})$, $\boldsymbol{R}_t$ is the compression ratio vector at time $t$, and $\boldsymbol{K}_{\mathrm{gc}}$ and $\boldsymbol{\alpha}$ are the constant coefficient vectors. Constrains (38)-(40) describe the operation characteristics of gas storages [32], where $\boldsymbol{E}_{\mathrm{st},t}^{\mathrm{w}}$ is the working gas volume; $\boldsymbol{S}_{\mathrm{st,dis,t}}$ is the gas discharging rate; $\boldsymbol{S}_{\mathrm{st,cha},t}^{\max}$ and $\boldsymbol{S}_{\mathrm{st,dis},t}^{\max}$ are the vectors of maximum gas charging/discharging rate at time $t$, respectively; and $\boldsymbol{K}_{\mathrm{st},1}$, $\boldsymbol{K}_{\mathrm{st},2}$, and $\boldsymbol{K}_{\mathrm{st},3}$ are the constant coefficient vectors. Constraint (41) ensures the gas balance at each node in each time period.

*2) Inequality Constraints of Gas Subsystem:*

$$-\boldsymbol{S}_{\mathrm{p}}^{\mathrm{c}}\leq\boldsymbol{S}_{\mathrm{p},t}\leq\boldsymbol{S}_{\mathrm{p}}^{\mathrm{c}} \quad (42)$$

$$\boldsymbol{R}^{\min}\leq\boldsymbol{R}_t\leq\boldsymbol{R}^{\max} \quad (43)$$

$$\boldsymbol{S}_{\mathrm{gc}}^{\min}\leq\boldsymbol{S}_{\mathrm{gc},t}\leq\boldsymbol{S}_{\mathrm{gc}}^{\max} \quad (44)$$

$$\boldsymbol{E}_{\mathrm{st}}^{\min}\leq\boldsymbol{E}_{\mathrm{st,t}}^{\mathrm{w}}\leq\boldsymbol{E}_{\mathrm{st}}^{\max} \quad (45)$$

$$-\boldsymbol{S}_{\mathrm{st,cha},t}^{\max}\leq\boldsymbol{S}_{\mathrm{st,dis},t}\leq\boldsymbol{S}_{\mathrm{st,dis},t}^{\max} \quad (46)$$

$$\boldsymbol{S}_{\mathrm{gw}}^{\min}\leq\boldsymbol{S}_{\mathrm{gw},t}\leq\boldsymbol{S}_{\mathrm{gw}}^{\max} \quad (47)$$

$$\boldsymbol{\Pi}_t^{\min}\leq\boldsymbol{\Pi}_t\leq\boldsymbol{\Pi}_t^{\max} \quad (48)$$

$$0 \le \boldsymbol{S}_{\mathrm{Lc},t} \le \boldsymbol{S}_{\mathrm{L},t} \tag{49}$$

where constraint (42) is the lower and upper bounds of the gas flows. Constrains (43)-(44) give the ranges of compression ratio and energy flow for gas compressors. Constrains (45)-(46) describe the limit of working gas volume and gas discharging rate for gas storages. Constrain (47) specifies the gas production limit of the gas wells, and constraint (48) limits the lower and upper bounds of the gas pressure. The range of gas load shedding is given in constrain (49).

*3) Equality Constraints of Power Subsystem:*

$$\boldsymbol{T}_t \text{-} \left(\boldsymbol{M}_{\mathrm{TL}}^{e}\right)^{\mathrm{T}} \boldsymbol{\theta}_t ./ \boldsymbol{X}_{\mathrm{T}} = 0 \tag{50}$$

$$\begin{aligned} &\boldsymbol{M}_{\mathrm{g}}^{\mathrm{e}} \boldsymbol{P}_{\mathrm{g},t} - \boldsymbol{M}_{\mathrm{L}}^{\mathrm{e}} (\boldsymbol{P}_{\mathrm{L},t} - \boldsymbol{P}_{\mathrm{Lc},t}) - \boldsymbol{M}_{\mathrm{TL}}^{\mathrm{e}} \boldsymbol{T}_t + \boldsymbol{M}_{\mathrm{gpg}}^{\mathrm{e}} \boldsymbol{P}_{\mathrm{gpg},t} \\ &\quad - \boldsymbol{M}_{\mathrm{p2g}}^{\mathrm{e}} \boldsymbol{P}_{\mathrm{p2g},t} + \boldsymbol{M}_{\mathrm{w}}^{\mathrm{e}} (\boldsymbol{P}_{\mathrm{w},t} - \boldsymbol{P}_{\mathrm{wc},t}) = 0 \end{aligned} \tag{51}$$

where constraints (50)-(51) describe the power flows of transmission lines and the power balance at electrical nodes, $\boldsymbol{T}_t$ is the vector of power flows at time $t$, $\boldsymbol{\theta}_t$ is the vector of phase angles at electrical nodes at time $t$, $\boldsymbol{X}_{\mathrm{T}}$ is the vector of reactance of transmission lines.

*4) Inequality Constraints of Power Subsystem:*

$$\boldsymbol{P}_{\mathrm{g}}^{\min} \le \boldsymbol{P}_{\mathrm{g},t} \le \boldsymbol{P}_{\mathrm{g}}^{\max} \tag{52}$$

$$-\boldsymbol{T}^{max} \le \boldsymbol{T}_t \le \boldsymbol{T}^{max} \tag{53}$$

$$0 \le \boldsymbol{P}_{\mathrm{Lc},t} \le \boldsymbol{P}_{\mathrm{L},t} \tag{54}$$

$$0 \le \boldsymbol{P}_{\mathrm{wc},t} \le \boldsymbol{P}_{\mathrm{w},t} \tag{55}$$

where constraints (52)-(53) describe the generation bounds of generators and transmission lines limiting capacities, respectively. Constraints (54)-(55) indicate the range of power load shedding and wind power curtailment, respectively.

*5) Equality Constraints of Coupling Facilities:*

$$\boldsymbol{P}_{\mathrm{gpg},t} - \boldsymbol{\eta}_{\mathrm{gpg}} \cdot \boldsymbol{S}_{\mathrm{gpg},t} = 0 \tag{56}$$

$$\boldsymbol{S}_{\mathrm{p2g},t} - \boldsymbol{\eta}_{\mathrm{p2g}} \cdot \boldsymbol{P}_{\mathrm{p2g},t} = 0 \tag{57}$$

where constraint (56) quantitates the relationship between the natural gas requirement $\boldsymbol{S}_{\mathrm{gpg},t}$ and the power generation output $\boldsymbol{P}_{\mathrm{gpg},t}$ for GPGs, in which $\boldsymbol{\eta}_{\mathrm{gpg}}$ is an efficiency coefficient vector. Constraint (57) quantitates the relationship between the power consumption $\boldsymbol{P}_{\mathrm{p2g},t}$ and natural gas production $\boldsymbol{S}_{\mathrm{p2g},t}$ for P2Gs, in which $\boldsymbol{\eta}_{\mathrm{p2g}}$ is an efficiency coefficient vector.

*6) Inequality Constraints of Coupling Facilities:*

$$\boldsymbol{P}_{\mathrm{gpg}}^{\min} \le \boldsymbol{P}_{\mathrm{gpg},t} \le \boldsymbol{P}_{\mathrm{gpg}}^{\max} \tag{58}$$

$$\boldsymbol{S}_{\mathrm{p2g}}^{\min} \le \boldsymbol{S}_{\mathrm{p2g},t} \le \boldsymbol{S}_{\mathrm{p2g}}^{\max} \tag{59}$$

where constraints (58)-(59) give the lower and upper bounds for $\boldsymbol{P}_{\mathrm{gpg},t}$ and $\boldsymbol{S}_{\mathrm{p2g},t}$, respectively.

*7) Energy Flow Model And Optimal Energy Flow*

In the risk evaluation of EGIES, the energy flow calculation for each sampled system state is carried out to determine whether the operational constraints are violated. The energy flow equations of EGIES consist of all the equality constraints mentioned above, which can be solved by the Newton method. If there are any violated constraint in the solutions of energy flow calculation, an optimal energy flow model is conducted to determine the minimum loss required to recover the EGIES from an abnormal state to the normal state. The optimal energy flow model of EGIES consists of an objective function with all the equality and inequality constraints (33)-(59) mentioned above. The minimized objective is the weighted sum of loss of gas loads $\boldsymbol{S}_{\mathrm{Lc},t}$, loss of electricity loads $\boldsymbol{P}_{\mathrm{Lc},t}$ and spilled renewable wind power $\boldsymbol{P}_{\mathrm{wc},t}$:

$$\min\ \omega_1 \sum_{i=1}^{N_{\mathrm{G}}} S_{\mathrm{Lc},t,i} + \omega_2 \sum_{i=1}^{N_{\mathrm{E}}} P_{\mathrm{Lc},t,i} + \omega_3 \sum_{i=1}^{N_{\mathrm{W}}} P_{\mathrm{wc},t,i} \tag{60}$$

where $\omega_1$, $\omega_2$ and $\omega_3$ are the weight coefficients. A primal-dual interior point algorithm [33] is used to solve the optimal model.

*C. Risk Evaluation Framework of EGIES Considering Pipeline Leakage Failure Modes*

*1) Risk Indices for EGIES*

Various risk indices for EGIES are systematically evaluated to indicate the risk level of EGIES. The risk indices are defined in terms of the six aspects: general load (electrical load and gas load) shedding, electrical load shedding, gas load shedding, wind power curtailment, minor gas leak and serious gas leak. There are three risk indices for each aspect. Take the risk indices based on general load shedding as an example:

$$\mathrm{LOLP} = \lim_{N\to\infty} \frac{1}{N} \sum_{i=1}^{N} \mathrm{Lolp}_i = \lim_{N\to\infty} \frac{1}{N} \sum_{i=1}^{N} \left( \frac{1}{8760} \sum_{t=1}^{8760} I_i^P(t) \right) \tag{61}$$

$$\mathrm{EENS} = \lim_{N\to\infty} \frac{1}{N} \sum_{i=1}^{N} \mathrm{Eens}_i = \lim_{N\to\infty} \frac{1}{N} \sum_{i=1}^{N} \sum_{t=1}^{8760} I_i^E(t) \tag{62}$$

$$\mathrm{LOLF} = \lim_{N\to\infty} \frac{1}{N} \sum_{i=1}^{N} \mathrm{Lolf}_i = \lim_{N\to\infty} \frac{1}{N} \sum_{i=1}^{N} \sum_{t=2}^{8760} I_i^F(t-1,t) \tag{63}$$

where LOLP (loss of general load probability) is equals to the sum of occurrence probability of random failure events where general load shedding occurs; $\mathrm{Lolp}_i$ denotes the total duration of general load shedding in the $i$th simulation year, and $I_i^P(t)$ is an index function which is equal to 1 if the general load shedding occurs at time $t$ in the $i$th simulation year, or 0 if otherwise. EENS means the expected general energy not supplied, $\mathrm{Eens}_i$ denotes the total general energy not supplied in the $i$th simulation year, and $I_i^E(t)$ is a function which is equal to the general energy not supplied at time $t$ in the $i$th simulation year. LOLF means the loss of general load frequency, $\mathrm{Lolf}_i$ denotes the number of occurrence of general load shedding in the $i$th simulation year, and $I_i^F(t\text{-}1,t)$ is a function which is equal to 1 if $I_i^P(t\text{-}1)=0$ and $I_i^P(t)=1$, otherwise it is equal to 0.

Similarly, the other risk indices for EGIES can be easily defined, including the $\mathrm{LOLP^e}$, $\mathrm{EENS^e}$, and $\mathrm{LOLF^e}$ based on electrical load shedding, the $\mathrm{LOLP^g}$, $\mathrm{EENS^g}$, and $\mathrm{LOLF^g}$ based on gas load shedding, the POWC (probability of wind power curtailment), EOWC (expectation of wind energy curtailment), and FOWC (frequency of wind power curtailment) based on wind power curtailments, the POML (probability of minor gas leak), EOML (expectation of minor gas leak), and FOML (frequency of minor gas leak) based on minor gas leaks, and the POSL (probability of serious gas leak), EOSL (expectation of serious gas leak), and FOSL (frequency of serious gas leak) based on serious gas leaks.

The stopping criterion for the risk evaluation is the coefficient of variation of EENS, which is calculated by

$$\beta = \mathrm{std(EENS)/EENS} \tag{64}$$

where std(EENS) is the standard deviation of EENS.

*2) Procedure of Risk Evaluation of EGIES*

The risk evaluation of EGIES considering the pipeline leakage failure modes consists of the following steps:

Step 1) Obtain the parameters of EGIES, including the yearly load curve, the transition rates of each component, and so on. Assume that all components are initially in the up state.

Step 2) In the given time span (one year), obtain the chronological pipeline state transition process for each pipeline by using the proposed bi-level Monte Carlo sampling method (see Section III above), and obtain the chronological component state transition processes for other components by using the traditional state duration sampling method.

Step 3) Obtain the chronological system state transition process by combining the chronological component state transition processes of all components.

Step 4) Conduct the system analysis for each different system state in the given time span (one year), in which the energy flow calculations are made to judge whether it is necessary to do an optimal load shedding calculation based on the model mentioned above (see Subsection B).

Step 5) Update the yearly risk indices according the results of system analysis. Check whether the stopping criterion meets the requirement. If not, go to Step 2); otherwise, output the final risk indices.

## V. Case Study

### A. Test System

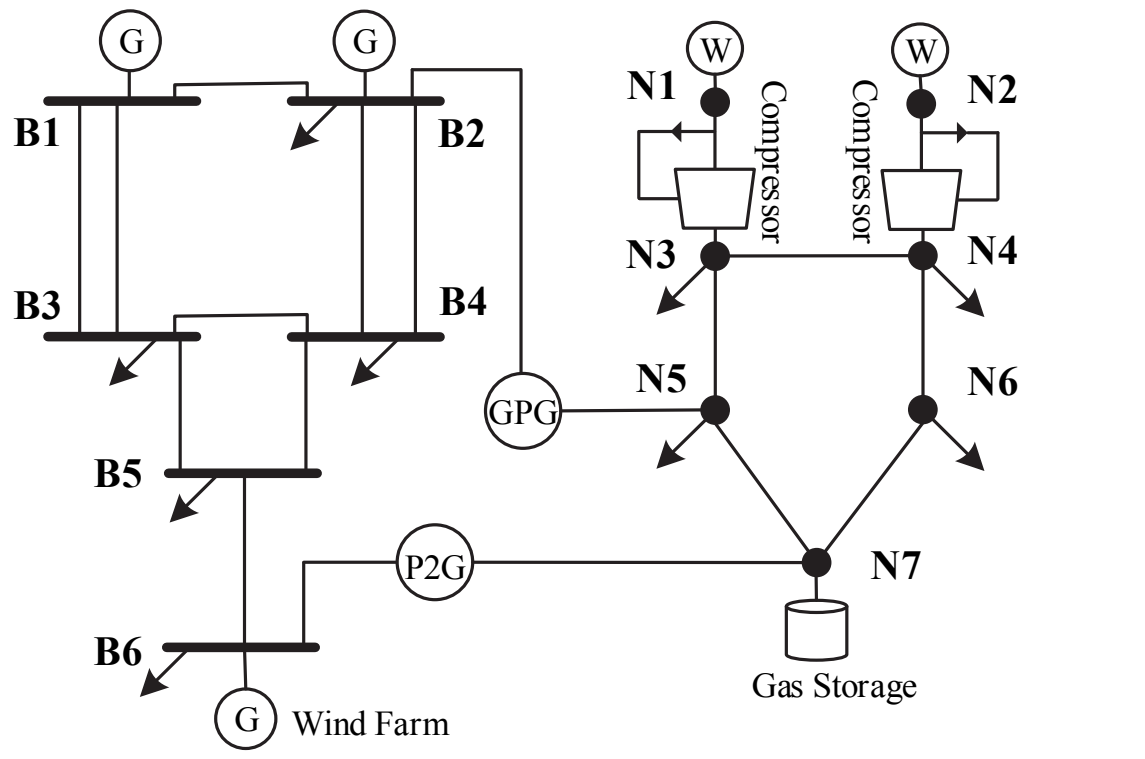

Fig. 5. An EGIES with a modified RBTS and a 7-node natural gas system

An EGIES with a modified RBTS and a 7-node natural gas system, which is shown in Fig.5, is applied to the proposed risk evaluation method for demonstrating the impacts of pipeline leakage failure modes.

To achieve a good balance between the power subsystem and gas subsystem in the EGIES, the capacity of generators and the system load level of RBTS [34] are increased to 1200MW of installed generation capacity and 925 MW of peak load, respectively. The gas network has a total 1200MW of gas generation capacity and a peak gas load with 800MW.

A 100MW generator at B2 in the RBTS is replaced by a 100MW GPG that is connected to node N5 in the gas network. A 150MW wind farm and a 100MW P2G are added at node B6. It is assumed that the 4 gas loads in the gas subsystem have the same daily load profiles as the electrical loads in the RBTS [34].

For the two pipeline leakage failure modes, the mean value of diameter of pinhole is estimated to be 20 mm with a standard deviation of 5 mm, and the mean value of diameter of hole is estimate to be 70 mm with a standard deviation of 10 mm. The other parameters for this test system are available in [35]. $\beta$=6% is specified as the convergence criterion.

### B. Natural Gas Pipeline Model

The input data of the four-state Markov model of pipeline are given in Table I, in which the frequencies of encountering different failures are taken from [26]. The three unknown transition rates $\lambda_{12}$, $\lambda_{13}$ and $\lambda_{14}$ in the model can be calculated using Eq. (7). Take a pipeline with the length of 100km for example, the transition rates $\lambda_{12}$, $\lambda_{13}$ and $\lambda_{14}$ are $1.103\times10^{-2}$, $3.310\times10^{-2}$, and $1.204\times10^{-2}$ occ./yr, respectively.

TABLE I
Pipeline Markov Model Parameters

| Frequency | Value (occ./(kkm·yr)) | Transition Rate | Value (occ./yr) |
|---|---|---|---|
| $f_0$ | 0.11 | $\lambda_{21}$ | 52.14 |
| $f_3$ | 0.33 | $\lambda_{32}$ | 12.17 |
| $f_4$ | 0.23 | $\lambda_{34}$ | 6.08 |
| \ | \ | $\lambda_{42}$ | 4380 |

In order to demonstrate the impacts of pipeline leakage failure modes on the risk evaluation of EGIES, a typical two-state model of pipeline without considering leakage failures is used for a comparison, which is shown in Fig. 6. In this two-state model, the leakage failures of pipeline are ignored, and only the rupture failure of pipeline is considered to be a down state. The parameter $\lambda^t_{21}$ is assumed to be equal to $\lambda_{21}$ in the four-state model, i.e., the repair process is considered to be the same and independent of pipeline failures. The transition rate $\lambda^t_{12}$ can be easily calculated by using the frequency-duration method [7] from the known failure frequency $f_0$ and the repair rate $\lambda^t_{21}$. For a pipeline with the length of 100km, the failure rate $\lambda^t_{12}$ is $1.100\times10^{-2}$ occ./yr.

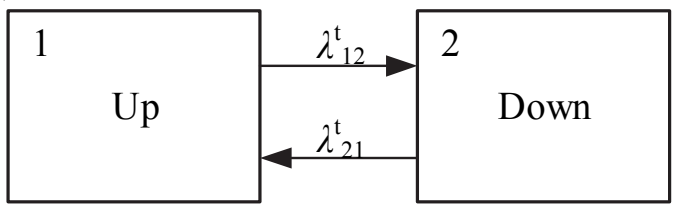

Fig. 6. Two-state model for a pipeline without considering leakage modes

### C. Risk Evaluation Analysis

The risk evaluation results for the two cases are shown in Table II. The definitions of the risk indices can be found in Section IV-C. In Case 1, the two-state model for all pipelines were considered and in Case 2, the proposed four-state Markov model for all pipelines were considered. It can be seen that the system indexes for incorporating pipeline leakage failures are much larger than those for only considering rupture failures. The differences due to leakage failures of the five pipelines are considerable in the results. This is because the pipelines are the main transmission links in the gas subsystem. This suggests that for an electricity-gas integrated energy system, if leakage failures are ignored in modeling as a usual assumption in current literatures, the risk indexes will be significantly underestimated and this will most likely lead to a misleading conclusion in system planning. Compared with the system indexes and power subsystem indexes, the gas subsystem indexes have much more differences between the two cases.

This obviously is due to the fact that the failures of gas pipelines have larger influences in the gas subsystem. Fig. 7 shows the differences between the two cases in the system EENS index and gas subsystem $EENS^g$ index over years in the simulation process.

TABLE II
RISK INDEXES OF THE TEST SYSTEM UNDER DIFFERENT PIPELINE MODELS

| Index | Case 1 | Case 2 | Difference |
|---|---|---|---|
| LOLP | $1.586\times10^{-3}$ | $1.944\times10^{-3}$ | 22.5% |
| LOLF(occ./yr) | 2.324 | 2.736 | 17.7% |
| EENS(MWh/yr) | $6.848\times10^{2}$ | $8.324\times10^{2}$ | 21.5% |
| $LOLP^e$ | $7.196\times10^{-4}$ | $7.382\times10^{-4}$ | 2.6% |
| $LOLF^e$(occ./yr) | 1.260 | 1.293 | 2.6% |
| $EENS^e$(MWh/yr) | $2.574\times10^{2}$ | $2.615\times10^{2}$ | 1.6% |
| $LOLP^g$ | $8.706\times10^{-4}$ | $1.209\times10^{-3}$ | 38.8% |
| $LOLF^g$(occ./yr) | 1.072 | 1.451 | 35.3% |
| $EENS^g$(MWh/yr) | $4.274\times10^{2}$ | $5.709\times10^{2}$ | 33.6% |
| POWC | $2.961\times10^{-4}$ | $3.218\times10^{-4}$ | 8.7% |
| FOWC(occ./yr) | 1.008 | 1.057 | 4.8% |
| EOWC(MWh/yr) | $1.204\times10^{1}$ | $1.255\times10^{1}$ | 4.2% |

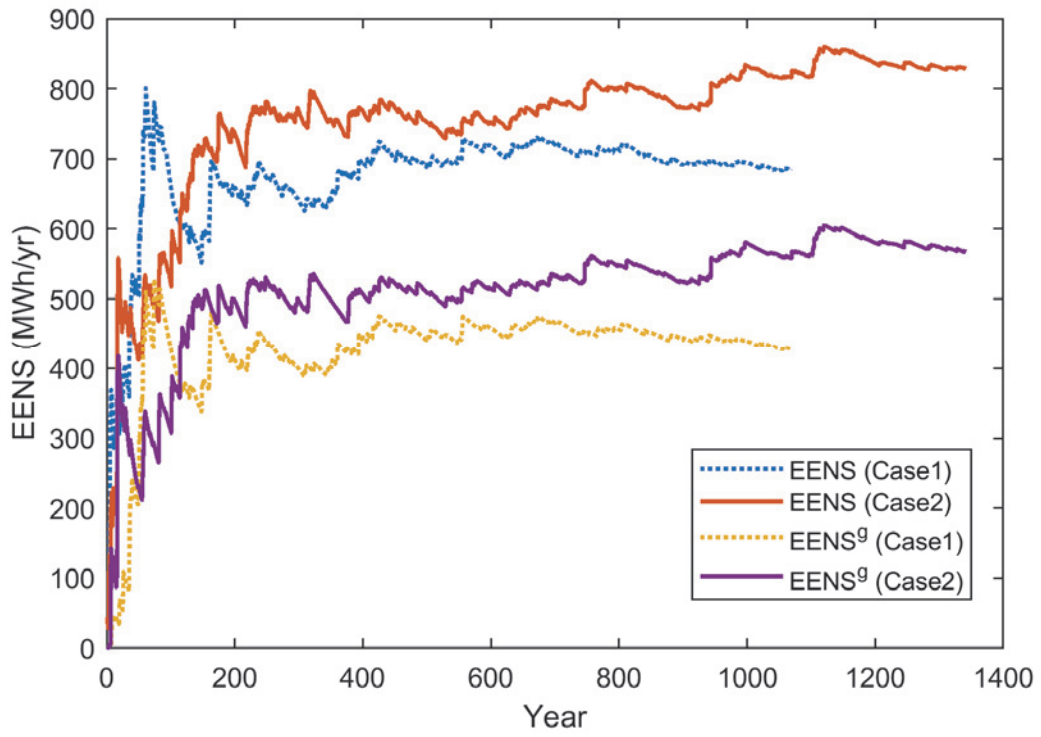

Fig. 7. Risk indexes EENS and $EENS^g$ of the test system for Cases 1 and 2

TABLE III
RESULTS AND COMPARISON OF GAS LEAK RISK INDEXES

| Index | Value | Index | Value | Ratio |
|---|---|---|---|---|
| LOLP | $1.944\times10^{-3}$ | POML | $9.295\times10^{-3}$ | 4.78 |
| LOLF | 2.736 | FOML | $1.948\times10^{-1}$ | 0.07 |
| EENS | $8.324\times10^{2}$ | EOML | $5.384\times10^{2}$ | 0.65 |
| POSL | $2.973\times10^{-5}$ | POML | $9.295\times10^{-3}$ | 312.65 |
| FOSL | $1.231\times10^{-1}$ | FOML | $1.948\times10^{-1}$ | 1.58 |
| EOSL | $2.054\times10^{1}$ | EOML | $5.384\times10^{2}$ | 26.21 |

The results and comparisons of gas leak risk indexes are given in Table III. LOLP, LOLF and EENS are the indices representing the curtailment risk of system general loads (both electrical and gas loads); POML, FOML and EOML are the indices for minor gas leakage risks and POSL, FOSL and EOSL are the indices for severe gas leakage risks. The definitions of all the indices can be found in Section IV-C. It can be seen from the first three rows in Table III that the POML index is approximately five times as large as the system LOLP index, the FOML index is lower than the system LOLF index, and the EOML is larger than the half of the system EENS index. This suggests that the gas leakage risk is comparable to or even possibly larger than the whole system load curtailment risk. Beyond the significance of the comparison against the system indices, the gas leakage indices can also provide the quantitative information on the safety and security risks associated with possible fires, explosions, and poisoning. It can be seen from the last three rows in Table III that for this test system, the frequency of serious leaks is comparable to that of minor leaks, whereas the probability and magnitude based indexes for serious leaks are much smaller than those for minor leaks. This indicates that even minor leaks cannot be ignored in the risk evaluation and that minor and serious leakage accidents should be both modeled as two different states as built in the proposed four-state model.

### D. *Impact of Failure Frequency*

As well known, there always exist quite large uncertainties in any failure statistics data. The frequencies of pinhole, hole and rupture accidents used in Table I may be low [26]. Two more cases with an increase of 50% in the failure frequencies are examined on the same test system. The system risk indexes are shown in Table IV. In Case 3, the two-state model for all pipelines without pipeline leakage modes were considered and in Case 4, the proposed four-state Markov model for all pipelines with pipeline leakage modes were considered. It can be seen that the difference for each system index increases approximately by 2 times if compared to that in Table II. This suggests that for an EGIES with more frequent failures in pipelines, ignoring the leakage failure modes will result in much larger underestimation errors in system risk indexes.

TABLE IV
SYSTEM RISK INDEXES UNDER DIFFERENT PIPELINE MODELS

| Index | Case 3 | Case 4 | Difference |
|---|---|---|---|
| LOLP | $1.573\times10^{-3}$ | $2.403\times10^{-3}$ | 52.8% |
| LOLF(occ./yr) | 2.276 | 3.177 | 39.6% |
| EENS(MWh/yr) | $7.058\times10^{2}$ | $9.921\times10^{2}$ | 40.6% |

TABLE V
GAS LEAK RISK INDEXES UNDER DIFFERENT CASES

| Index | Case 2 | Case 4 | Difference |
|---|---|---|---|
| POML | $9.295\times10^{-3}$ | $1.540\times10^{-2}$ | 65.7% |
| FOML(occ./yr) | $1.948\times10^{-1}$ | $2.703\times10^{-1}$ | 38.8% |
| EOML(MWh/yr) | $5.384\times10^{2}$ | $8.553\times10^{2}$ | 58.9% |
| POSL | $2.973\times10^{-5}$ | $4.721\times10^{-5}$ | 58.8% |
| FOSL(occ./yr) | $1.231\times10^{-1}$ | $2.076\times10^{-1}$ | 68.6% |
| EOSL(MWh/yr) | $2.054\times10^{1}$ | $3.250\times10^{1}$ | 58.2% |

The gas leakage risk indexes in Case 2 and Case 4 were summarized in Table V. It can be seen that the gas leak indexes in Case 4 are much larger than those in Case 2. The majority of percentage differences in the gas leakage indices between the two cases are larger than the increase of 50% in the failure frequencies of pipelines (input data). This suggests that the gas leakage risk in the EGEIS is expected to increase non-linearly with the failure frequencies of pipelines.

## VI. CONCLUSIONS

This paper presents a method to incorporate pipeline leakage failures in the EGIES risk evaluation. It includes the development of a Markov state transition model of pipeline for representing the multi-state and multi-failure-mode transition process and a bi-level Monte Carlo sampling method in the risk evaluation. Additionally, a stochastic topology change based network model of EGIES considering the pipeline leakage failure modes is presented. The risk indices for EGIES based on load shedding minimization, including those specifically for gas leakage risks, are also proposed.

An EGIES with a modified RBTS and a 7-node gas system, which represents a regional integrated electricity-gas system, was used as an example to demonstrate an application of the proposed method and models. The results indicate that pipeline leakage failures have significant impacts on the risk of EGIES, particularly for an EGIES with more frequent failures in pipelines. Ignoring pipeline leakage failures in the risk evaluation of EGIES will result in an overly underestimation of system risk and most likely a misleading conclusion in system planning.